\documentclass[twocolumn,english,prl,showpacs]{revtex4}
\usepackage[T1]{fontenc}
\usepackage[latin9]{inputenc}
\usepackage{amssymb}
\usepackage{graphicx}
\usepackage{amsmath,color}
\usepackage{mathrsfs}
\usepackage{float}
\usepackage{indentfirst}

\makeatletter

\def\journal #1, #2, #3, 1#4#5#6{{\sl #1~}{\bf #2}, #3 (1#4#5#6) }

\newcommand{\COMMENTED}[1]{}

\makeatother

\begin{document}

\title{Pole expansion of self-energy and interaction effect on topological insulators}

\author{Lei Wang$^{1,2*}$, Hua Jiang$^{2}$, Xi Dai$^1$ and X. C. Xie$^{2}$}

\affiliation{$^{1}$Beijing National Lab for Condensed Matter Physics and Institute of Physics, Chinese Academy of Sciences, Beijing 100190, China}

\affiliation{$^{2}$International Center for Quantum Materials, Peking University, Beijing 100871, China}

\affiliation{*Current address: Theoretische Physik, ETH Zurich, 8093 Zurich, Switzerland}


\begin{abstract}
We study effect of interactions on time-reversal-invariant topological insulators. Their topological indices are expressed by interacting Green's functions. Under the local self-energy approximation, we connect topological index and surface states of an \emph{interacting} system to an auxiliary \emph{noninteracting} system, whose Hamiltonian is related to the pole-expansions of the local self-energy. This finding greatly simplifies the calculation of interacting topological indices and gives an noninteracting pictorial description of interaction driven topological phase transitions. Our results also bridge studies of the correlated topological insulating materials with the practical dynamical-mean-field-theory calculations.
\end{abstract}






\maketitle

\textit{Introduction--}Recently, there has been a rising interest on a class of topological phase of matter: the topological insulators (TI), see \cite{Hasan:2010p23520, Moore:2010p15238} for reviews. Questions like what is the interaction effect on the TI and how to characterize the topological phase beyond single particle basis pose intriguing challenges. There are some previous studies on the interacting effect on TIs. \citet{Wang:2010p25724} performed exact diagonalization and quantum Monte Carlo studies of the interacting Haldane model. \citet{Hohenadler:2011p28789,Zheng:2011p43686,Yu:2011p35101,Wu:2011p36258} performed quantum Monte Carlo, variational cluster approximation and cluster dynamical-mean-field-theory (DMFT) studies on the Kane-Mele-Hubbard (KMH) model.  One interesting finding in these studies is that the topological phases could give their way to a gapped featureless state (spin liquid phase for KMH model) at small value of the spin orbital coupling. It is possible that the fluctuation effect driven by interaction breaks the topological phase significantly. This scenario goes beyond the mean-field treatments \cite{Rachel:2010p20458, Wang:2010p25724}. Since in the latter case, long range orders (LRO) are usually needed to compete with the topological phases.

However, one unsatisfactory feature in these studies is that there lacks a direct characterization of the topological order: the phase boundaries are mostly determined by indirect signatures such as the correlation functions for LRO or the closing and reopening features of excitation gaps, rather than a sharp change of one topological index. This leaves the mechanism and the nature of the correlation induced topological transitions unclear.

In this Letter, we study the interaction effects on TI based on their single-particle Green's functions. we connect topological index and surface states of an \emph{interacting} system to the indices of an auxiliary \emph{noninteracting} system, whose Hamiltonian is related to the pole-expansions of the local self-energy. This finding greatly simplifies the calculation of interacting topological indices and gives an noninteracting pictorial description of interaction driven topological phase transitions.

The system's topological index $n_{G}$ is determined by the interacting Green's function through \cite{Qi:2008p12545,Wang:2010p25171}:

\begin{eqnarray}
n_{G} & =  & \frac{\pi^2 }{15} \varepsilon_{\mu \nu \rho\sigma\tau } \mathrm{Tr} \int  \frac{\mathrm{d}^{4}k \mathrm{d}\omega}{(2\pi)^{5}} \, \nonumber \\
  & &G \partial_{\mu} G^{- 1} G \partial_{\nu} G^{- 1} G \partial_{\rho} G^{- 1}G \partial_{\sigma} G^{- 1}G \partial_{\tau} G^{- 1}
\label{eqn:ishikawa}
\end{eqnarray}
where $G$ is Matsubara Green's function \footnote{The system should be gapped, otherwise integrating out fermions is invalid and in general there is no quantized topological index.}, $\varepsilon_{\mu\nu\rho\sigma\tau}$ is the five-order anti-symmetric tensor, $\mu, \nu...\tau$ indices denote the frequency-momentums $(\omega, k_x, k_y, k_{z}, k_{\lambda})$. The same formula applies to the Chern number and Z$_{2}$ index. In latter case, $k_{\lambda}$ is a pumping parameter extends three dimensional TI to four dimension\cite{Wang:2010p25171}. Physically, Chern number describes quantum Hall conductance, Z$_{2}$ index describes magnetoelectric response of an insulator \cite{Hasan:2010p23520}. The insulator's Z$_{2}$ index should be ether $0$ or $1$ if there is time-reversal symmetry. In that case, only the parity of $n_{G}$ is meaningful.
Formula Eqn.\ref{eqn:ishikawa} involves only the single particle Green's function and allows analysis of the effect of interaction, disorder and finite temperature on equal footing. In \cite{Wang:2011p43509}, the authors point out the frequency-domain-winding-number (FDWN) could change the topological index without developing long range orders. However, the paper made one crude assumption (besides the local self-energy assumptions) that the self-energy is diagonal in the orbital basis and has no orbital dependence. In this paper, we relax this assumption by adopting the pole-expansion (PE) for general local self-energies. We link the topological indices and surface states of an \emph{interacting} system to an \emph{noninteracting} system thereof.

\textit{Pole expansion of local self-energy--} Considering a TI with noninteracting Hamiltonian $H_{\bold k}$, who is a matrix of the size $m$. Assume that the local interaction generates a self-energy without momentum dependence except in the Hartree-Fock part, thus, the self-energy can be written into the PE form as ${\Sigma}(\omega,\bold k)={\Sigma}_{\bold k}+{V}^{\dagger}(i\omega-{P})^{-1} {V}$.
This type of self-energy can be obtained by the DMFT, which is believed to be reliable for 3D correlated systems \cite{Georges:1996p5571}. ${\Sigma}_{\bold k}$ is a $m\times m$ matrix denotes the asymptotic (Hartree-Fock) contributions to the self-energy. ${P}$ and $V$ are frequency-independent matrices of the size $ N_{P}\times N_{P}$ and $N_{P}\times m$ respectively. $P$ and $V$ determine the position and weight of poles of the self-energy. Physically, PE corresponds to represent the interaction induced local self-energy to coupling with noninteracting baths. One could introduce a noninteracting pseudo-Hamiltonian of the form \cite{Savrasov:2006p3872} $
\tilde{H}_{\bold k} = \left(\begin{array}{cc}
  H_{\bold{k}}+{\Sigma}_{\bold k} -\mu& {V}^{\dagger} \\
  {V} & {P}
\end{array}\right) $, which has the matrix size $N= N_{P}+m$. The corresponding Green's function reads


\begin{widetext}


\begin{eqnarray}
\tilde{G} = (i\omega-\tilde{H}_{\bold{k}})^{-1}
  = \left(\begin{array}{cc} i\omega+\mu-H_{\bold{k}}-{\Sigma}_{\bold k} & -{V}^{\dagger} \nonumber \\  -{V} & i\omega-{P} \end{array}\right)^{-1}
 = \left(\begin{array}{cc} [ i\omega +\mu- H_{\bold{k}} - {\Sigma}_{\bold k}-{V}^{\dagger}(i\omega-{P})^{-1} {V}]^{-1} & \ldots \\ \vdots & \ddots \end{array}\right)
\end{eqnarray}

Clearly, $\tilde{G}$ is related to the interacting Green's function $G= [i\omega+\mu-H_{\bold{k}}-\Sigma(\omega)]^{-1}$ through a projection $G =\mathcal{P}^{\dagger} \tilde{G}
\mathcal{P} $. The projection operator $\mathcal{P}= \left(\begin{array}{cccc}
  1 & 0 & \ldots \\
  0 & 1 & \ldots \\
  \vdots & \vdots & \ddots
 \end{array}\right)$ is a $N\times m$ matrix projects to the upper-left elements of $\tilde{G}$. Notice that the integrand of Eqn.\ref{eqn:ishikawa} is:

\begin{eqnarray}
  & &  \mathrm{Tr} G \partial_{\omega} G^{-1}G\partial_{k_x} G^{- 1} G
  \partial_{k_y} G^{- 1} G
  \partial_{k_z} G^{- 1} G
  \partial_{k_\lambda} G^{- 1}  \nonumber\\
 & = &-\mathrm{Tr}  \frac{\partial G}{\partial \omega} \partial_{k_x} G^{- 1} G \partial_{k_y} G^{- 1} G \partial_{k_z} G^{- 1}  G \partial_{k_\lambda} G^{- 1}   \nonumber\\
  & =&- \mathrm{Tr} \mathcal{P}^{\dagger} \frac{\partial \tilde{G}}  {\partial \omega} \mathcal{P}\mathcal{P}^{\dagger} \partial_{k_{x}} \tilde{G}^{-1} \mathcal{P}  \mathcal{P}^{\dagger}\tilde{G}\mathcal{P}
\mathcal{P}^{\dagger} \partial_{k_{y}} \tilde{G}^{-1} \mathcal{P}    \mathcal{P}^{\dagger}\tilde{G}\mathcal{P}
  \mathcal{P}^{\dagger} \partial_{k_{z}} \tilde{G}^{-1} \mathcal{P}   \mathcal{P}^{\dagger}\tilde{G}\mathcal{P}
 \mathcal{P}^{\dagger} \partial_{k_{\lambda}} \tilde{G}^{-1} \mathcal{P}    \nonumber\\
  & = &- \mathrm{Tr}  \frac{\partial \tilde{G}}{\partial \omega} \mathcal{P}\mathcal{P}^{\dagger} \partial_{k_x}
  \tilde{G}^{- 1} \mathcal{P}\mathcal{P}^{\dagger}  \tilde{G} \mathcal{P}\mathcal{P}^{\dagger}  \partial_{k_y} \tilde{G}^{- 1} \mathcal{P}\mathcal{P}^{\dagger} \tilde{G} \mathcal{P}\mathcal{P}^{\dagger}  \partial_{k_z} \tilde{G}^{- 1} \mathcal{P}\mathcal{P}^{\dagger} \tilde{G} \mathcal{P}\mathcal{P}^{\dagger}  \partial_{k_\lambda} \tilde{G}^{- 1} \mathcal{P}\mathcal{P}^{\dagger} \nonumber\\
  & = & -\mathrm{Tr} \frac{\partial \tilde{G}}{\partial \omega} \partial_{k_x}
  \tilde{G}^{- 1} \tilde{G} \partial_{k_y} \tilde{G}^{- 1} \tilde{G} \partial_{k_z} \tilde{G}^{- 1} \tilde{G} \partial_{k_\lambda} \tilde{G}^{- 1} \nonumber\\
  & =& \mathrm{Tr} \tilde{G} \partial_{\omega} \tilde{G}^{-1}\tilde{G}\partial_{k_x} \tilde{G}^{- 1} \tilde{G}
  \partial_{k_y} \tilde{G}^{- 1} \tilde{G}
  \partial_{k_z} \tilde{G}^{- 1} \tilde{G}
  \partial_{k_\lambda} \tilde{G}^{- 1}
   \label{eqn:proof}
 \end{eqnarray}
\end{widetext}
where we have used the identities $\partial_{k_{i}}G^{-1}=-\partial_{k_{i}}(H_{\bold k}+\Sigma_{\bold k} ) = \mathcal{P}^{\dagger} \partial_{k_{i}} \tilde{G}^{-1} \mathcal{P} $ and $ \mathcal{P}\mathcal{P}^{\dagger} \partial_{k_{i}}\tilde{G}^{-1} \mathcal{P}\mathcal{P}^{\dagger}=\partial_{k_{i}}\tilde{G}^{-1}$. Base on Eqn.\ref{eqn:proof}, one can proof that the topological (Chern or Z$_{2}$) indices calculated from $G$ and $\tilde{G}$ are identical, \textit{i.e.},
\begin{equation}
n_{G} = n_{\tilde{G}}
\label{eqn:central}
\end{equation}

Further more, the surface and bulk Green's functions are related through the projection $G^{\mathrm{surf}} = \mathcal{Q}^{\dagger} G \mathcal{Q}$. $\mathcal{Q}$ consists of a Fourier transformation of (let's say) momentum $k_z$ to real space, then project to the surface norms to $z$-axis. Since the two projection operators $\mathcal{P}$ and $\mathcal{Q}$ act in different spaces, one has $[ \mathcal{P}, \mathcal{Q}] = 0$. Thus,

\begin{equation}
 G^{\mathrm{surf}}=
   \mathcal{Q}^{\dagger} \mathcal{P}^{\dagger} \tilde{G} \mathcal{P}
   \mathcal{Q} = \mathcal{\mathcal{P}^{\dagger} Q}^{\dagger} \tilde{G}\mathcal{Q} P = \mathcal{\mathcal{P}^{\dagger}} \tilde{G}^{\mathrm{surf}} \mathcal{P}
   \label{eqn:surface}
\end{equation}
which means that the surface Green's function of an \emph{interacting} system ($G^{\mathrm{surf}}$) is related to an auxiliary \emph{noninteracting} system's surface Green's function ($ \tilde{G}^{\mathrm{surf}}$) through the same projection $\mathcal{P}$.

Eqn.\ref{eqn:central} and \ref{eqn:surface} are the central results of the paper. They link the topological index and surface states of an \emph{interacting} model to a \emph{noninteracting} model. The latter is purely determined by the \emph{noninteracting} pseudo-Hamiltonian $\tilde{H}_{\bold k}$. For a given local self-energy, one could construct its corresponding pseudo-Hamiltonian and calculate its topological index fellowing conventional procedures for noninteracting systems \cite{Fukui:2005p38594,Fukui:2007p38592,Essin:2009p26094}. This fact greatly reduces the computational complexity of calculating topological indices of interacting systems. Moreover, the relationship Eqn.\ref{eqn:central} and Eqn.\ref{eqn:surface} provide a way of understanding interaction induced topological transition by its noninteracting counterpart.

\begin{figure}[tbp]
\centering
  \includegraphics[width=8cm]{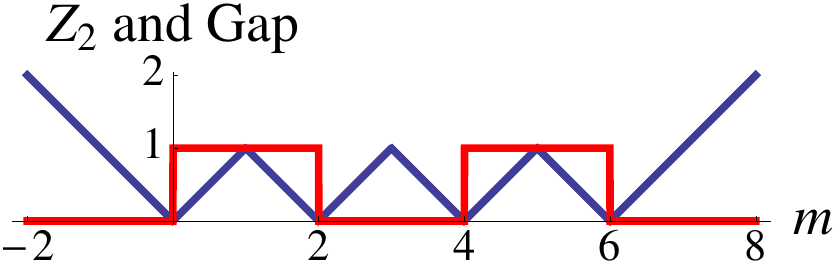}
\caption{(Color online) Z$_{2}$ index (red line) and gap (blue line) of noninteracting model (Eqn.\ref{eqn:noninteracting}) as a function of $m$. The phase boundaries between different Z$_{2}$ indices are in accordance with the gap closing points.}
\label{fig:phasediag}
\end{figure}

\textit{Applications--} We demonstrate how the general rules Eqn.\ref{eqn:central} and Eqn.\ref{eqn:surface} applies to an concrete model for interacting Z$_{2}$ insulator. We choose noninteracting Hamiltonian to be \footnote[1]{we set $\Gamma^{1}=\sigma^{x}\otimes\tau^{x}$, $\Gamma^{2}=\sigma^{y}\otimes\tau^{x}$, $\Gamma^{3}=\sigma^{z}\otimes\tau^{x}$, $\Gamma^{4}=\mathbb{I}_{2}\otimes\tau^{y}$,$\Gamma^{5}=\mathbb{I}_{2}\otimes\tau^{z}$.},
\begin{equation}
H_{\bold k}=\sin k_{x}\Gamma^{1}+\sin k_{y}\Gamma^{2}+\sin k_{z}\Gamma^{3}+\mathcal{M}({\bold k})\Gamma^{5}
\label{eqn:noninteracting}
\end{equation}
where $\mathcal{M}({\bold k})=m-3+(\cos k_{x}+\cos k_{y}+\cos k_{z})$. The Hamiltonian describes a 3D (strong) Z$_{2}$ TI for $0<m<2$ and  $4<m<6$. Inside the TI phase the gap of the system $\Delta$ scales linearly with $m$, see Fig.\ref{fig:phasediag}. We set $m=1$ in the following studies. Notice that the eigenvalues of $H_{\bold k}$ are $\pm E_{\bold k}$, with $ E_{\bold k} = \sqrt{\sin^{2} k_{x}+\sin^{2} k_{y} +\sin^{2} k_{z}+\mathcal{M}^{2}(\bold k)}$. It also determines the band gap $\Delta=1$ and band top $\Gamma=5$.



In the following we consider the effect of local interaction on the TI phase. Rather than trying to solve the interacting model explicitly (leave to future studies), we set up a few ansatz of the self-energies (see Table.\ref{tab:selfen}) and calculate its interacting Z$_{2}$ index using the following methods:
(1) We get the interacting Z$_{2}$ number based on its FDWN \cite{Wang:2011p43509};
(2) With the PE coefficients of the given self-energy (see Table\ref{tab:selfen}) we construct the pseudo-Hamiltonian $\tilde{H}_{\bold k}$ and calculate its Z$_{2}$ index fellowing noninteracting procedures \cite{Fukui:2007p38592}. We also study the gap closing condition and surface states of the interacting system through the noninteracting reference system $\tilde{H}_{\bold k}$.


\begin{table*}
\caption{Self-energies and their PE coefficients. Interacting Z$_{2}$ indices calculated by two methods (see text) are in accordance with each other.}
\label{tab:selfen}
\tabcolsep 0pt
\vspace{5pt}
\def\temptablewidth{\textwidth}
{\rule{\temptablewidth}{1pt}}
\begin{tabular*}{\temptablewidth}{@{\extracolsep{\fill}}cccccccc}
Cases & ${\Sigma}(\omega,\bold k)$& $N_{P}$ & $\Sigma_{\bold k}$ & ${V}^{\dagger}$ &  ${P}$ & Z$_{2}=1$ & Z$_{2}=0$ \\ \hline
 (a) &  $\frac{V^{2}}{i\omega-P}$ & $4$ & $0$  & $V \mathbb{I}_{4}$       & $P \mathbb{I}_{4}$  & $|P|\Delta > V^{2}$ & $|P|\Gamma<V^{2}$ \\
 (b) & $\frac{V^{2}}{i\omega+P}+\frac{V^{2}}{i\omega-P}$  & $8$ & $0$ & $(V,V)\otimes \mathbb{I}_{4}$ & $diag(P,-P)\otimes \mathbb{I}_{4}$  & $P\neq0$ & $P=0$ \\

\end{tabular*}
    {\rule{\temptablewidth}{1pt}}
\end{table*}

\begin{figure}[tbp]
\centering
    \includegraphics[width=5cm,height=4cm]{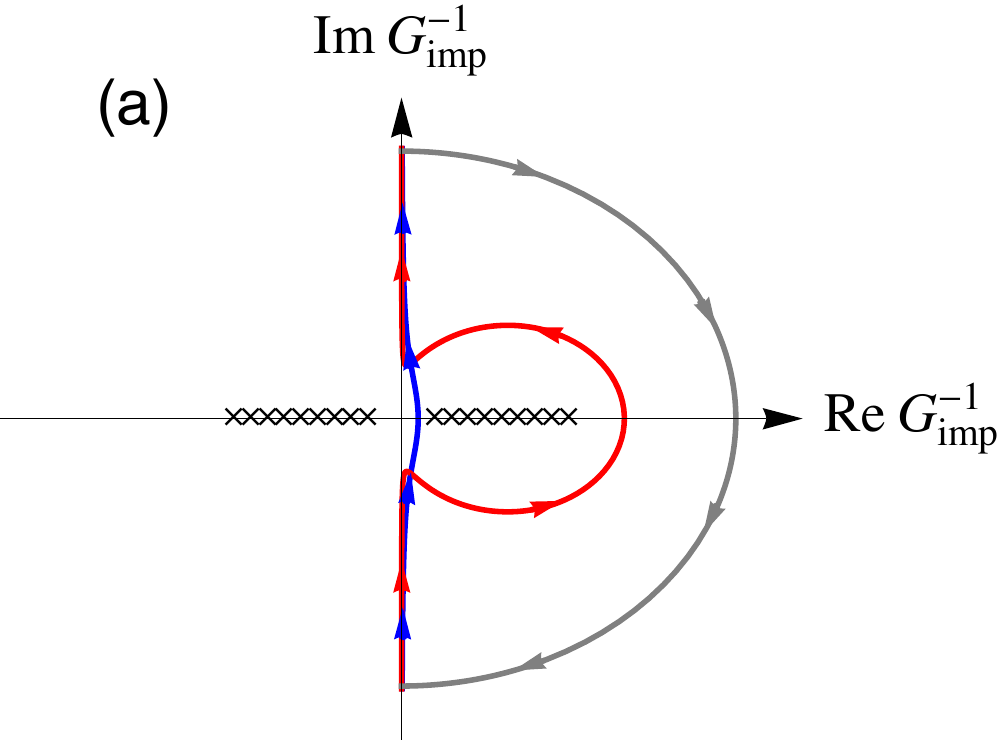}
    \includegraphics[width=3cm,height=4cm]{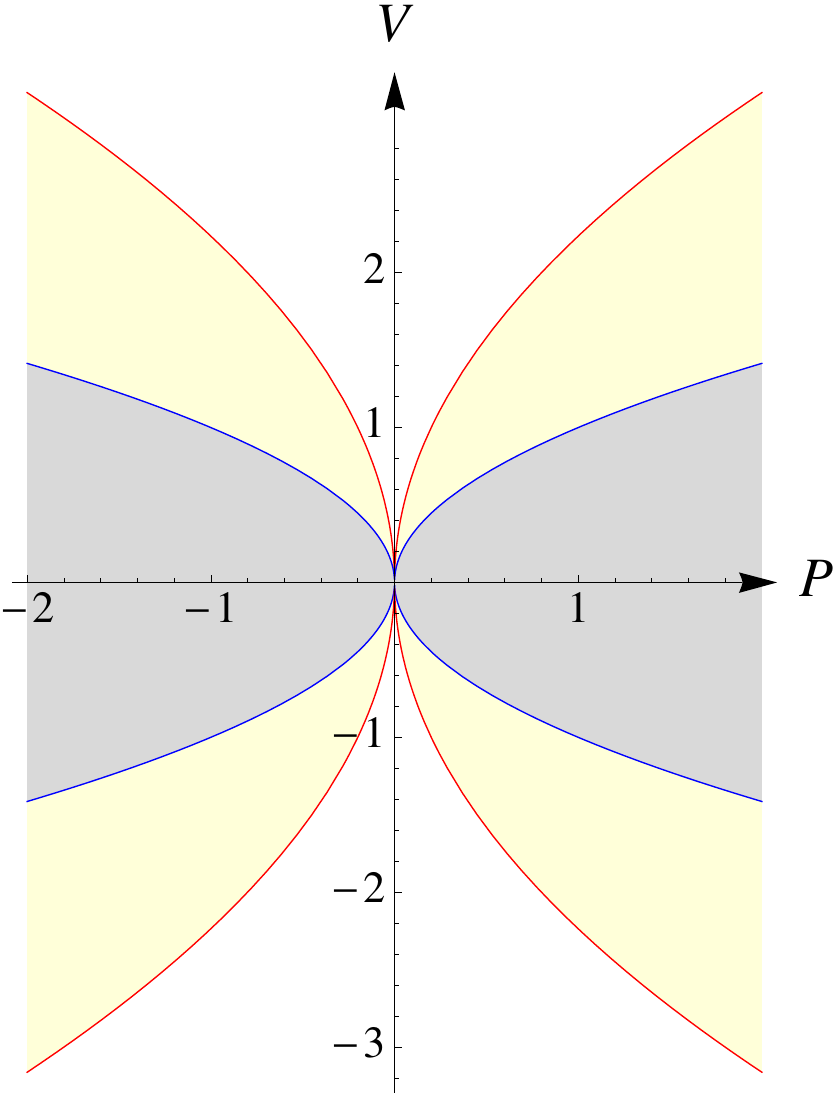}

\caption{(Color online) (Left panel) Paths of the mapping $\omega\mapsto G_{\mathrm{imp}}^{-1}=i\omega - \frac{V^{2}}{i\omega-P}$ with $V=1$, $P=2$ (blue line) and $P=0.15$ (red line). Gray line indicates the large circle we appended to close the contour. Crosses on real axis indicate the eigenenergy continuum of the noninteracting Hamiltonian, whose lower (upper) bound equals to $\Delta$ ($\Gamma$). Topological transition between TI and metal occurs when $|P|\Delta=V^{2}$, where the blue line touches the lower bound. The gap reopens for $|P|\Gamma<V^{2}$, results into a normal insulating phase. (Right panel) Phase diagram determined by calculating the Z$_{2}$ index of the pseudo-Hamiltonian (see the PE coefficients in Table.\ref{tab:selfen}). Shaded and white region have Z$_{2}=1$ and $0$ respectively, the yellow region denotes the metallic state. The phase boundaries $|P|\Delta=V^{2}$ and $|P|\Gamma=V^{2}$  are in accordance with the FDWN analysis.}
\label{fig:contour}
\end{figure}

We start with the case (a): $\Sigma(\omega,\bold k) = \frac{V^{2}}{i\omega-P}$. 
We draw the path defined by $\omega\mapsto G_{\mathrm{imp}}^{-1}(\omega)=i\omega-\frac{V^{2}}{i\omega-P}$ on the complex plane and find that the topological transition point corresponds to where the path touch the lower edge of the eigenenergy continuum, \textit{i.e.} the band gap $\Delta$. This defines the phase boundary between the TI and metal phase $\frac{V^{2}}{P}=\pm\Delta$, see Fig.\ref{fig:contour}(a). We then calculate the Z$_{2}$ index of the noninteracting pseudo-Hamiltonian $\tilde{H}_{\bold k}$ with its single-particle engenstates
\footnote{To do the calculation, we need to build up the pseudo time reversal operator as $\tilde{\mathcal{T}} = diag(1,1,...,1) \otimes\mathcal{T} $ with $\mathcal{T} = \Theta \mathcal{K}$. $\Theta=i\sigma_{y}\otimes \mathbb{I}_{2}$ and $\mathcal{K}$ is the complex conjugate operator.  Actually for this particular case, Z$_{2}$ index could also be calculated by simply parity counting \cite{Fu:2007p3992} (one need also extend the definition of the parity operator). However, since in general self-energy may broken the inversion symmetry. Thus we using the more general integration algorithm \cite{Fukui:2007p38592} to calculate the Z$_{2}$ index.}. The resulting phase diagram is shown in Fig.\ref{fig:contour}(b). Phase boundaries are in accordance with the prediction based on FDWN.

\begin{figure}[!t]
\centering
    \includegraphics[width=8cm]{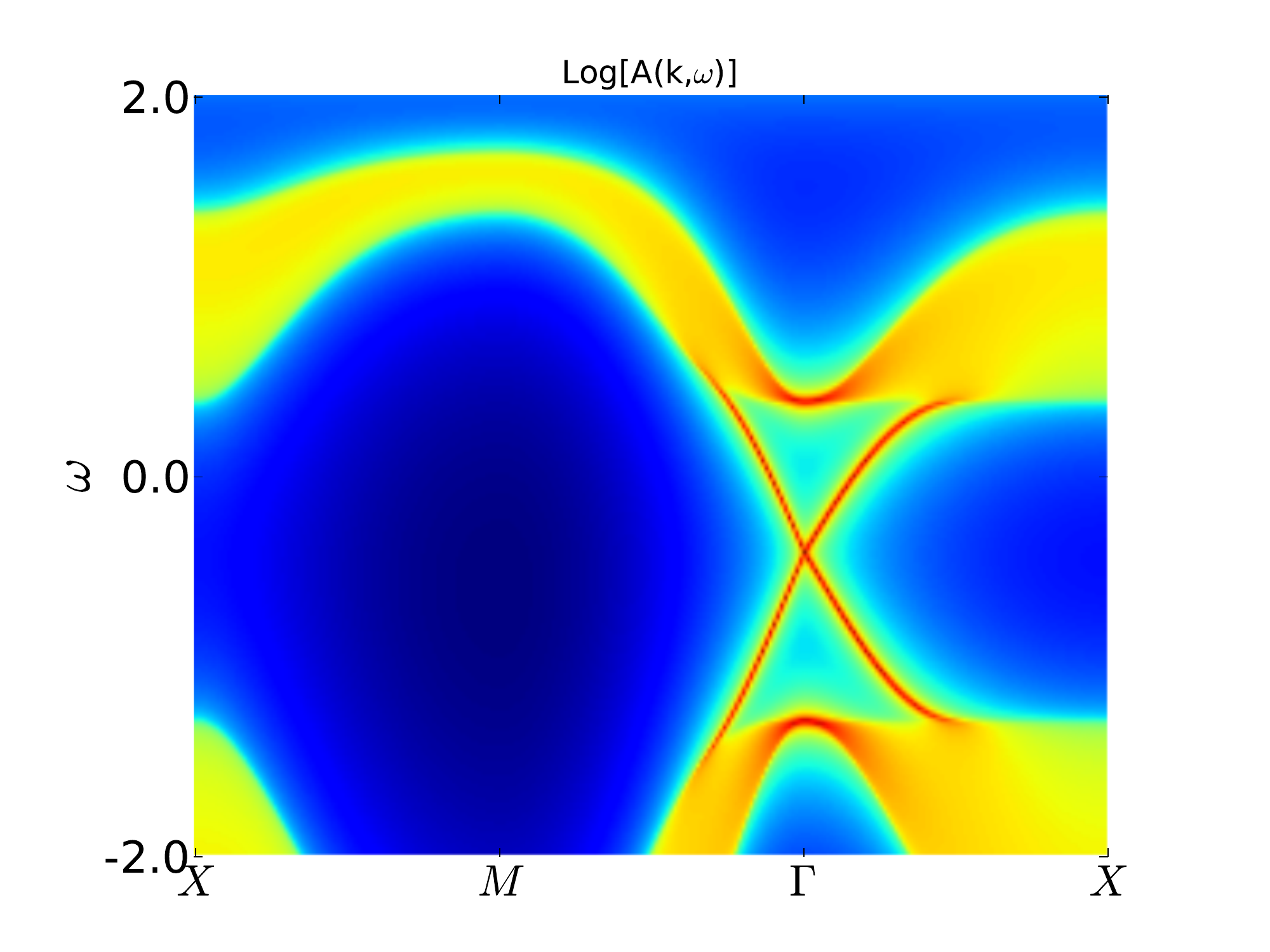}
      \includegraphics[width=8cm]{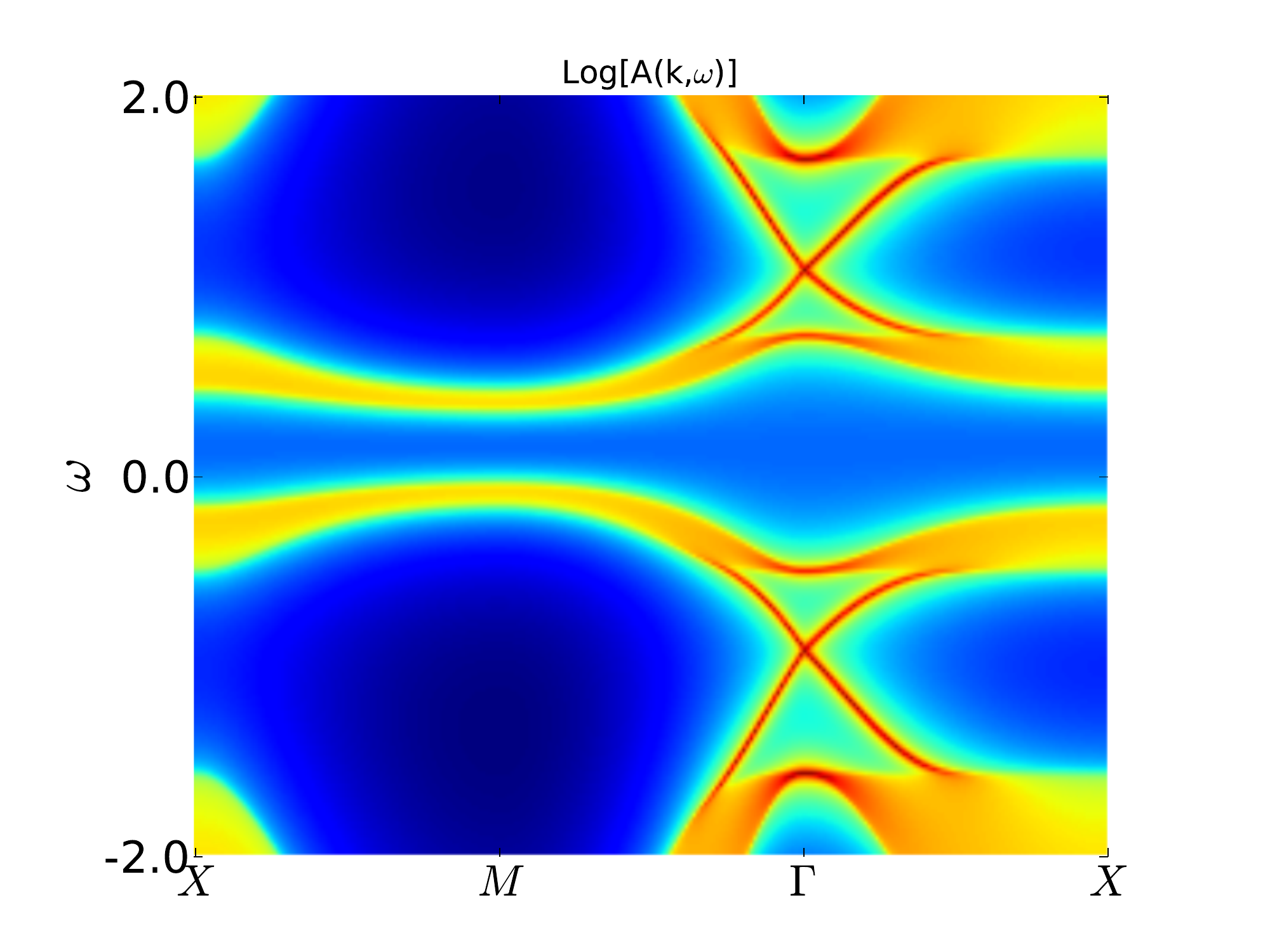}

\caption{Surface spectral functions for self-energy case (a) with $V=1$. (upper panel) $P=2$ is in TI phase, (lower panel) $P=0.15$ is normal insulator. The chemical potential lies at zero. }
\label{fig:surface}
\end{figure}

The surface Green's function could be directly calculated for the interacting system with given self-energy or from the surface Green's function associated to the pseudo-Hamiltonian $\tilde{H}_{\bold k}$, (Eqn.\ref{eqn:surface}). We have numerically verified that the two approaches give identical results. The resulting surface spectral functions are shown in Fig.\ref{fig:surface}. There are gapless surface states crossing the Fermi surface when the system is in the interacting topological phase, Fig.\ref{fig:surface} upper panel.

We then move to the case (b): $\Sigma(\omega,\bold k) = \frac{V^{2}}{i\omega+P}+\frac{V^{2}}{i\omega-P}$ \footnote{This ansatz captures the Mott transition scenario of the ``two-site DMFT''\cite{Potthoff:2001p37308}}. It has been shown that the interacting system is gapful and its Z$_{2}$ index equals to $0$ for $P=0$ and $1$ otherwise\cite{Wang:2011p43509}. Diagonalizing $\tilde{H}_{\bold k}$ shows that the noninteracting system is gapful for $P\neq0$, with Z$_{2}$ index equal to $1$, which validates the aforementioned result. The gap of the pseudo-Hamiltonian closes at $P=0$. At first look, this seems to be inconsistent with the Eqn.\ref{eqn:central} because the ground state of $\tilde{H}_{\bold k}$ is gapless while the interacting system is gapful. This paradox is solved by noticing that for $P=0$ the two poles of case (b) merge into one but we are still constructing the pseudo-Hamiltonian with $N_{P}=8$. When treat this particular point with $N_{P}=4$, case (b) reduced to case (a) with $P=0$, which is topological trivial and consistent with the FDWN prediction. Lesson from this example is that when constructing the pseudo-Hamiltonian, degenerate poles should be treated carefully.


%
%

\textit{Discussions--} Direct dealing with the frequency-momentum integration in Eqn.\ref{eqn:ishikawa} is the most general way of calculating interacting topological index. It does not make any priori assumptions on the form of the self-energies. However, as been pointed out in \cite{Wang:2011p43509}, the distribution of the integrand in Eqn.\ref{eqn:ishikawa} is very inhomogeneous. Numerical integration for inversion symmetric Z$_{2}$ TIs is even harder, since all of the topological information concentrate around those time-reversal invariant momenta \cite{Fu:2007p3992}. Integrate with Monte Carlo technique or adaptive meshes may overcome the difficulties to some degree.  On the other hand, the concept of FDWN greatly simplify the calculation and gives an appealing picture for the interaction effect on TIs.  However, the usefulness of FDWN is only limited to the case where the self-energy is diagonal in the orbital basis (besides the local assumptions). The PE approach lies in the middle of the two aforementioned limiting cases. It relaxes the orbital-independent assumption and is more general for higher dimensional TIs. PE is merely the discrete approximation for the self-energy, which provides a systematic way to approach the true self-energy by making the number of poles denser and denser.

Under the local self-energy approximation, the PE approach could also be used to calculate the Chern and Z$_{2}$ indices for two dimensional systems. However, the main advantage of PE approach lies in high dimensions. For two dimensional systems, ground state wave function (GSWF) under twisted boundaries could be used to extract topological index\cite{Lee:2008p17479}. However, GSWF is very hard to get for higher dimensional interacting fermionic models. In fact, comparing to Green's functions, GSWF contains redundant information for topological properties. Furthermore, there are large number of analytical or numerical approaches of getting interacting Green's function for correlated fermionic models.

Base on these analysis, we anticipate the local self-energy approximation and PE technique would of great use in subsequent studies of 3D interacting TIs.

\textit{Conclusion--} We show that the local self-energy contains the topological signature of interacting TIs. With the help of pole-expansion of the local self-energy, we find that the topological index of an \emph{interacting} topological (Chern or Z$_{2}$) insulator is identical to the one for an \emph{noninteracting} model. This findings greatly simplify the calculation of the topological index for interacting TIs and give an appealing noninteracting picture for interaction induced topological phase transitions. Techniques report here could have greatly potential use in direct determination of the topological indices in the realistic LDA+DMFT studies of the 3D correlated TI materials.

\textit{Note--} See Ref.\cite{Wang:2012p50499} for a recent application of pole-expansion method on Bernevig-Hughes-Zhang model. There is an independent application of FDWN and pole-expansion idea to topological quantum phase transitions \cite{Budich:2012p51247}.

\textit{Acknowledgment--}The work is supported by NSF-China and MOST-China.
We thank Yuan Wan, X.-L Qi, Cenke Xu and Y. Hatsugai for helpful discussions.

\bibliography{/Users/wanglei/Documents/Papers/papers}
\end{document}